\documentclass[12pt]{article}
\usepackage{amsfonts,latexsym,amsmath,latexsym,amssymb}
 
\newtheorem{definition}{Definition}

\def\C{{\mathbb{C}}}

\def\N{{\mathbb{N}}}

\newcommand{\sL}[0]{\mathrm{\mathbf{L}}}

\newcommand{\cH}{{\cal H}}

\newcommand{\ket}[1]{|#1\rangle}
\newcommand{\bra}[1]{\langle #1|}

\title{The $2$-local Hamiltonian problem\\ encompasses NP}

\author{Pawel Wocjan\thanks{e-mail: {\protect\tt
\{wocjan,eiss\_office\}@ira.uka.de}}\, and Thomas Beth \\ \small
Institut f{\"u}r Algorithmen und Kognitive Systeme, Universit{\"a}t
Karlsruhe,\\[-1ex] \small Am Fasanengarten 5, D-76\,131 Karlsruhe,
Germany}

\begin{document}

\maketitle

\abstract{We show that the NP complete problems MAX CUT and
INDEPENDENT SET can be formulated as the $2$-local Hamiltonian problem
as defined by Kitaev. He introduced the quantum complexity class BQNP
as the quantum analog of NP, and showed that the $5$-local Hamiltonian
problem is BQNP-complete. It is not known whether the $s$-local
Hamiltonian problem is BQNP-complete for $s$ smaller than
$5$. Therefore it is interesting to determine what problems can be
reduced to the $s$-local Hamiltonian problem. Kitaev showed that
$3$-SAT can be formulated as a $3$-local Hamiltonian problem. We
extend his result by showing that $2$-locality is sufficient in order
to encompass NP.}

\section{Introduction}
The field of complexity theory has long been studied in terms of
classical physics. One of the most important complexity categories is
the NP complexity class \cite{GJ:79}. With the beginning of quantum
computing there emerged quantum extensions of these complexity
categories. For instance, Kitaev introduced the quantum complexity
class BQNP as the quantum analog of NP, and showed that the $5$-local
Hamiltonian problem is BQNP complete \cite{KSV:02,AN:02}. It is not
known whether the $s$-local Hamiltonian problem is BQNP-complete for
$s=2,3,4$.

In this paper we study the complexity of the $2$-local Hamiltonian
problem in classical terms. We prove that NP problems are polynomially
reducible to the $2$-local Hamiltonian problem. This reduction is
shown by formulating the NP-complete problems MAX CUT and INDEPENDENT
SET as a $2$-local Hamiltonian problem.

The ideas of our proof are motived by results of statistical physics
on the complexity of computing the energy of ground states of spin
glasses. It is known that the problem of computing the energy of
ground states of (1) spin glasses with no exterior magnetic field and
of (2) planar spin glasses within a magnetic field are
NP-complete. These results are established by showing that the problem
of finding the energy of ground states is equivalent to solving the
NP-complete problem MAX CUT in (1) and INDEPENDENT SET in (2)
\cite{Barahona:82}.

\section{Local Hamiltonian problem}
We repeat the necessary definitions concerning the local Hamiltonian
problem. For the general context we refer the reader to
\cite{KSV:02,AN:02}.

Let $\cH:=\C^2$ denote the Hilbert space of a single qubit and
$\cH^{\otimes n}$ the joint Hilbert space of $n$
qubits. $\sL(\cH^{\otimes s})$ denotes the set of linear operators
from $\cH^{\otimes s}$ to $\cH^{\otimes s}$. Let $A\in\sL(\cH^{\otimes
s})$ be an arbitrary operator and $S\subseteq\{1,\ldots,n\}$ with
$|S|=s$. We denote by $A[S]\in\sL(\cH^{\otimes n})$ the embedding of
the operator $A$ into the Hilbert space $\cH^{\otimes n}$, i.e., the
operator that acts as $A$ on the qubits specified by $S$.
\begin{definition}[Local Hamiltonian]${}$\\
An operator $H:\cH^{\otimes n}\rightarrow\cH^{\otimes n}$ is called an
$s$-local Hamiltonian if it is expressible in the form
\begin{equation}
H=\sum_j H_j[S_j]\,,
\end{equation}
where each term $H_j\in\sL(\cH^{\otimes |S_j|})$ is a positive
semidefinite operator of bounded norm $\|H_j\|\le 1$ acting on a set
$S_j$, $|S_j|\le s$.
\end{definition}

\begin{definition}[The local Hamiltonian problem]${}$\\
Let $H$ be an $s$-local Hamiltonian and $a,b$ be nonnegative real
numbers, where $s$ is a constant and $b-a>n^{-\alpha}$ ($\alpha>0$ is
a constant). The $s$-local Hamiltonian problem is to determine if
either
\begin{enumerate}
\item $H$ has an eigenvalue not exceeding $a$, or
\item all eigenvalues of $H$ are greater than $b$.
\end{enumerate}
The defined problem is a promise problem: we know that one of the two
situations occurs.
\end{definition}

\section{Max cut}
We show that the MAX CUT problem may be formulated as the $2$-local
Hamiltonian problem defined by Kitaev.

The MAX CUT problem is defined as follows \cite{Karp:72,GJ:79}:
\begin{itemize}
\item INSTANCE: Weighted graph $G=(V,E)$, weight $w(e)\in\N$ for each
$e\in E$, positive integer $w$.
\item QUESTION: Is there a partition or {\em cut} $C=(V_0,V_1)$ of $V$
into disjoint $V_0$ and $V_1$ such that the sum of the weights of the
edges that have one end point in $V_0$ and one endpoint in $V_1$ is at
least $w$?
\end{itemize}
This problem remains NP-complete if $w(e)=1$ for all $e\in E$ (the
SIMPLE MAX CUT problem) \cite{GJS:76}.

In the following we consider the SIMPLE MAX CUT problem. Let $G=(V,E)$
a graph. To have a unique representation we denote edges as ordered
pairs $(k,l)$ with $k<l$. Following the idea of \cite{Barahona:82} we
associate a binary variable $X_k\in\{0,1\}$ to each vertex $k\in
V$. For an assignment $X_1,\ldots,X_n$ let us define $V_0:=\{k\mid
X_k=0\}$, and $V_1:=\{k\mid X_k=1\}$. This defines the cut
$C=(V_0,V_1)$ of $G$. Let us also define $E_{00}$, and $E_{11}$ as the
set of edges with both end points in $V_0$ and $V_1$,
respectively. The cut refers to the set of edges that cross between
``zero'' vertices, i.e.\ $k\in V_0$, to the ``one'' vertices, i.e.\
$k\in V_1$. Let $E_{01}$ be the set of edges in the cut, i.e., all
edges with the first vertex in $V_0$ and the second in $V_1$. $E_{10}$
is defined analogously. The weight of the cut is $w(C)=
|E_{01}|+|E_{10}|$.

Clearly, as $X_1,\ldots,X_n$ varies over all assignments, the
corresponding cut $C$ varies over all cuts of $G$. Especially, there
is a one-to-one correspondence between assignments and cuts.

There is a cut whose weight is at least $w$ if and only if there is an
assignment to the variables $X_1,\ldots,X_n$ such that
\begin{equation}\label{eq:maxCutIneq}
w(C)=|E_{01}|+|E_{10}|=\sum_{(k,l)\in E} (1-X_k) X_l + X_k(1-X_l)\ge w\,.
\end{equation}
To define our $2$-local Hamiltonian encoding the MAX CUT problem we
start from the inequality
\begin{eqnarray}
\sum_{(k,l)\in E} X_k X_l + (1-X_k)(1-X_l) & = & \label{eq:even} \\
\sum_{(k,l)\in E} \big(1-(1-X_k) X_l - X_k (1-X_l)\big) & \le &
|E|-w
\end{eqnarray}
that is obtained from inequality~(\ref{eq:maxCutIneq}) by multiplying
with $-1$, and adding $|E|$ on both sides. We use the $2$-local
projection
\begin{equation*}
P_{even} =\left(
\begin{array}{cccc}
1 & 0 & 0 & 0 \\ 0 & 0 & 0 & 0 \\ 0 & 0 & 0 & 0 \\ 0 & 0 & 0 & 1
\end{array}
\right)
= 
\ket{00}\bra{00}+\ket{11}\bra{11}
\end{equation*}
to define our $2$-local Hamiltonian such that its eigenvalues are
given by (\ref{eq:even}). The Hamiltonian is defined as
\begin{equation}
H = \sum_{(k,l)\in E} P_{even}[k,l]\,.
\end{equation}
Clearly, the eigenstates of $H$ are given by the computational basis
states. We denote them by
$\ket{X_1}\otimes\cdots\otimes\ket{X_n}$. The energy of the state
$\ket{X_1}\otimes\cdots\otimes\ket{X_n}$ is given by (\ref{eq:even}).

Therefore, $G$ has a cut whose weight is at least $w$ if and only if
$H$ has an eigenvalue that is smaller or equal to $|E|-w$. Since all
eigenvalues of $H$ are natural numbers, we may formulate this question
as the $2$-local Hamiltonian by choosing $a:=|E|-w+0.5$ and
$b:=a+0.25$. This choice ensures that either the first or the second
situation occurs in Definition~2. Furthermore, we have $b-a>1/n^2$ for
all $n\ge 3$.  Hence, we have shown that the SIMPLE MAX CUT problem
can be reduced to the $2$-local Hamiltonian problem defined by Kitaev.

\section{Independent set}
We show that the INDEPENDENT SET problem may be formulated as the
$2$-local Hamiltonian problem defined by Kitaev.

The INDEPENDENT SET problem is defined as follows \cite{GJ:79}:
\begin{itemize}
\item INSTANCE: Graph $G=(V,E)$, positive integer $v\le |V|$.
\item QUESTION: Does $G$ contain an independent set whose cardinality
is at least $v$, i.e., a subset $V'\subseteq V$ such that $|V'|\ge v$
and such that no two vertices in $V'$ are joined by an edge in $E$?
\end{itemize}
The INDEPENDENT SET problem remains NP-complete for cubic planar
graphs \cite{GJS:76}. A graph is called cubic if all vertices have
degree $3$, i.e., all vertices are connected to exactly three
vertices.

Following the idea of \cite{Barahona:82} we associate a variable
$X_k\in\{0,1\}$ to each vertex $k\in V$. There is an independent set
whose cardinality is at least $v$ if and only if there is an
assignment to the variables $\{X_k\mid k\in V\}$ such that
\begin{equation}\label{eq:independentSet}
\sum_{k\in V} X_k - \sum_{(k,l)\in E} X_k X_l\ge v\,.
\end{equation}
This is seen as follows. If $V'$ is an independent set whose
cardinality is at least $v$, then the assignment $X_k=1$ for $k\in V'$
and $X_k=0$ for $k\in V\setminus V'$ fulfills
inequality~(\ref{eq:independentSet}). 

Now let $X_1,\ldots,X_n$ be an assignment that fulfills
inequality~(\ref{eq:independentSet}). If $V'=\{k\mid X_k=1\}$ is not
an independent set, then we must have $|V'|\ge v+p$, where
$p:=\sum_{(k,l)\in E} X_k X_l>0$ is the ``penalty'' for $V'$ not being
an independent set. Let $(\tilde{k},\tilde{l})\in E$ with
$X_{\tilde{k}}=X_{\tilde{l}}=1$. By removing $\tilde{k}$ from $V'$
(i.e.\ setting $X_{\tilde{k}}:=0$) the cardinality of $V'$ drops by
$1$, while $p$ drops by at least $1$. After repeating this several
times, we end up with an independent set whose cardinality is at least
$v$.

To construct our $2$-local Hamiltonian encoding the INDEPENDENT SET
problem we start from the inequality
\begin{equation}\label{eq:equivalentIS}
\sum_{k\in V} (1-X_k) + \sum_{(k,l)\in E} X_k X_l\le |V|-v\,.
\end{equation}
that is equivalent to inequality~(\ref{eq:independentSet}). We use the
$1$-local projection
\begin{equation*}
P_0 = \left(\begin{array}{cc} 1 & 0 \\ 0 & 0\end{array}\right) =
\ket{0}\bra{0}
\end{equation*}
and the $2$-local projection
\begin{equation*}
P_{11} = 
\left(
\begin{array}{cccc}
0 & 0 & 0 & 0 \\
0 & 0 & 0 & 0 \\
0 & 0 & 0 & 0 \\
0 & 0 & 0 & 1
\end{array}
\right)
=\ket{11}\bra{11}\,
\end{equation*}
to define our $2$-local Hamiltonian such that its eigenvalues are
given by the left hand side of (\ref{eq:equivalentIS}). The
Hamiltonian is defined as
\begin{equation}\label{eq:IndepSetHam}
H = \sum_{k\in V} P_0[k] + \sum_{(k,l)\in E} P_{11}[k,l]\,.
\end{equation}
Clearly, the eigenstates of $H$ are given by the computational basis
states. We denote them by
$\ket{X_1}\otimes\cdots\otimes\ket{X_n}$. The energy of the state
$\ket{X_1}\otimes\cdots\otimes\ket{X_n}$ is the left hand side of
(\ref{eq:equivalentIS}). Therefore, $G$ has an independent set whose
cardinality is at least $v$ if and only if $H$ has an eigenvalue that
is smaller or equal to $|V|-v$. Since all eigenvalues of $H$ are
natural numbers, we may formulate this question as the $2$-local
Hamiltonian by choosing $a:=|V|-v+0.5$ and $b:=a+0.25$. This choice
ensures that either the first or the second situation occurs in
Definition~2. Furthermore, we have $b-a>1/n^2$ for all $n\ge 3$.
Hence, we have shown that the INDEPENDENT SET problem can be reduced
to the $2$-local Hamiltonian problem defined by Kitaev.

\section*{Acknowledgments}
This work was supported by grants of the BMBF project MARQUIS
01/BB01B. We would like to thank H.~Ros{\'e}, T.~Asselmeyer, and
A.~Schramm for interesting discussions.


\begin{thebibliography}{KSV02}

\bibitem[AN02]{AN:02}
D.~Aharonov and T.~Naveh.
\newblock {Quantum NP - A Survey}.
\newblock LANL e-print quant-ph/0210077, 2002.

\bibitem[Bar82]{Barahona:82}
F.~Barahona.
\newblock On the computational complexity of ising spin models.
\newblock {\em J. Phys. A: Math. Gen.}, 15:3241--3253, 1982.

\bibitem[GJ79]{GJ:79}
M.~R. Garey and D.~S. Johnson.
\newblock {\em {Computers and Intractability: A guide to the Theory of
  NP-Completeness}}.
\newblock W.~H.~Freeman and Company, 1979.

\bibitem[GJS76]{GJS:76}
M.~R. Garey, D.~S. Johnson, and L.~Stockmeyer.
\newblock {Some simplified NP-complete graph problems}.
\newblock {\em Theoretical Computer Science}, 1(2):237--267, 1976.

\bibitem[Kar72]{Karp:72}
R.~M. Karp.
\newblock Reducibility among combinatorial problems.
\newblock In R.~E. Miller and J.~W. Thatcher, editors, {\em {Complexity of
  Computer Computation}}, pages 85--103. Plenum Press, New York, 1972.

\bibitem[KSV02]{KSV:02}
A.~Yu. Kitaev, A.~H. Shen, and M.~N. Vyalyi.
\newblock {\em {Classical and Quantum Computation}}, volume~47.
\newblock American Mathematical Society, 2002.

\end{thebibliography}
\end{document}